\documentclass[aps,pra,amsmath,amssymb,showpacs,twocolumn]{revtex4-1}

\usepackage[T1]{fontenc}
\usepackage{amssymb,amsmath}
\usepackage{amsfonts}
\usepackage{graphicx}
\usepackage{amsthm}
\usepackage{color}
\usepackage{mathbbol}
\usepackage{epstopdf}
\usepackage{float}

\usepackage{tikz}
\usepackage{verbatim}

\def\vec#1{\mathbf{#1}}
\def\ket#1{|#1\rangle}

\def\bra#1{\langle#1|}

\def\tr{\mathrm{tr}}

\newcommand{\bn}{{\mathbf n}}

\newcommand{\floor}[1]{\lfloor #1 \rfloor}

\newcommand{\bJ}{{\bf J}}

\begin{document}

\title{Tensor eigenvalues and entanglement of symmetric states}
\author{F.~Bohnet-Waldraff$^{1,2}$, D.~Braun$^{1}$, and O.~Giraud$^{2}$}
\affiliation{$^{1}$Institut f\"ur theoretische Physik, Universit\"at T\"{u}bingen, 72076 T\"ubingen, Germany\\
$^{2}$\mbox{LPTMS, CNRS, Univ.~Paris-Sud, Universit\'e Paris-Saclay, 91405 Orsay, France}
}
\begin{abstract}
Tensor eigenvalues and eigenvectors have been introduced in the recent mathematical literature as a generalization of the usual matrix eigenvalues and eigenvectors.  We apply this formalism to a tensor that describes a multipartite symmetric state or a spin state, and we investigate to what extent the corresponding tensor eigenvalues contain information about the multipartite entanglement (or, equivalently, the classicality) of the state.  This extends previous results connecting entanglement to spectral properties related to the state. While for spin-1 states the positivity of the smallest tensor eigenvalue is equivalent to separability, we show that for higher values of the angular momentum there is a correlation between entanglement and the value of the smallest tensor eigenvalue.

\end{abstract}
\date{August 11, 2016}
\pacs{03.65.Aa, 03.65.Ca, 03.67.-a}
\maketitle

\section{Introduction}

In the study of multipartite entanglement, symmetric multipartite states have drawn some attention recently \cite{MultGeoEnt, DetecEntDicke, MaxEntGeoMeasure}.  One reason for that is that they span a Hilbert space whose dimension grows only linearly with the number of constituents, rather than exponentially for arbitrary multipartite states. They are therefore easier to deal with than generic states, and they provide a first step towards a more general understanding of multipartite entanglement. A pure symmetric $N$-qubit state can be written as a superposition of the Dicke states familiar in quantum optics. A Dicke state is a state of $N$ two-level atoms (i.e.~qubits) where a given number of excitations is symmetrically distributed over the $N$ constituents, so that the state is invariant under permutations of the qubits. Such states have important technological potential for quantum storage, as the coupling constants of photons to atoms can effectively be increased by a factor $\sqrt{N}$ when coupling the atoms symmetrically to the light field \cite{Dicke54}. Another physical realization of Dicke states is provided by angular momentum eigenstates, i.e.~spin-$j$ states arising as collective angular momentum states of $N=2j$ physical spins-1/2. The Dicke states are formally equivalent to eigenstates $\ket{j,m}$ of operators $\bJ^2$ and $J_z$, where $J_x,J_y,J_z$ are the usual angular momentum operators. A mixed symmetric state is then defined as a mixture of pure symmetric states (note that this notion is distinct from that of 'symmetrized mixed state', which would be a tensor product of spin-$1/2$ density matrices symmetrized by summing over all permutations).

Among the pure spin-$j$ states, spin coherent states (also called SU(2)-coherent states) are the ones that come as close as possible to the ideal of a classical phase space point, in the sense that their quantum fluctuations for the angular moment components are as small as allowed by Heisenberg's uncertainty relation \cite{Per72}. Furthermore, they keep this property under the dynamics induced by Hamiltonians linear in the angular momentum components, corresponding physically for example to precession in a magnetic field. For a spin-$j$ coherent state, the expectation value of the angular momentum operator in a specific direction $\bn$ is $\langle\bJ.\bn\rangle=\hbar j$, a feature not true for a general pure spin-$j$ state. In this sense a spin-$j$ coherent state points in a well-defined direction (note that all pure spin-1/2 states are coherent states, as they can be specified by a Bloch vector on the unit sphere). If a spin-$j$ coherent state is interpreted as a symmetric $N$-fold tensor product of $N=2j$ qubits, it can be expressed simply as the tensor product of $N$ identical spin-1/2 coherent states. Therefore spin-$j$ coherent states coincide with symmetric separable pure states. Classical spin-$j$ states are defined as statistical mixtures of spin coherent states \cite{hillery_nonclassical_1987,ClassSpinStates,mari_positive_2012}. When expressed in the Dicke basis, they can be seen as separable symmetric $N$-qubit states.

Just as entanglement of a quantum state can be measured as the distance to the set of separable states, the quantumness (or non-classicality) of a spin-$j$ state can be measured as its distance to the set of classical states \cite{QQQ}.  
Our purpose here is to investigate quantumness properties of a state from its spectral properties. There has been substantial research trying to figure out what entanglement properties can be derived from the spectrum of eigenvalues of the density matrix representing a composite system \cite{separability_prop,arunachalam_is_2014,gurvits_largest_2002,kus_geometry_2001,verstraete_maximally_2001,hildebrand_positive_2007,johnston_separability_2013}, and how to directly access the spectrum experimentally without having to reconstruct the full density matrix \cite{ekert_direct_2002,tanaka_determining_2014,ganguly_witness_2014}. Measures of entanglement based on the spectrum have the immediate advantages of being relatively easy to compute, and to be invariant under unitary transformations, i.e.~to capture ``absolute separability'' \cite{kus_geometry_2001}. Other well-known entanglement criteria are based on bounds of spin-correlations \cite{TothGuehne,UshPraRaj07}, which in turn exploit the positive-partial transpose (PPT) criterion. In \cite{PRL2015} we introduced a tensorial representation for spin states. In this representation, a spin-$j$ density matrix is expanded as a sum over matrices of dimensions $(2j+1)\times(2j+1)$, and the expansion coefficients take the form of a tensor $A_{\mu_1\mu_2\ldots\mu_{N}}$ with $N=2j$ indices. We showed in \cite{etaPaper} that the PPT criteria applied to symmetric multi-qubit states can be unified by means of a matrix $T$, obtained from the tensor representation of the equivalent spin-$j$ state by splitting the set of indices in two subsets, and considering each set as coding for the row or column index of the $T$-matrix. Positive partial transpose is then equivalent to positivity of the $T$-matrix, and correlation criteria for observables, such as spin-squeezing inequalities, can also be derived from positivity of $T$ \cite{etaPaper}.

In the light of these entanglement criteria based on spectral properties of the density matrix, or on positivity of the $T$-matrix constructed from the tensor $A$, one may wonder whether the spectrum of the tensor $A$ itself contains deeper information about the entanglement of the state. While the spectral theory of matrices is more than one century old, its extension to tensors is much more recent. The spectral theory of tensors has developed a lot in the past decade, and various tools have been proposed in the mathematical literature to tackle this problem (see \cite{RoughVersionQi} for a short review, and also Section \ref{SectionTensorEigevalues} below). But the relevance of the spectral theory of tensors for the separability (or classicality) problem has just recently attracted some attention in the quantum information community. For example in \cite{QiEnt16} it was shown that for pure states the largest tensor eigenvalue is equal to the geometric measure of entanglement, i.e.~the maximal overlap of the state with a pure separable state. This entanglement measure is in fact essentially equivalent to finding the best rank-one approximation of the tensor. Therefore, the largest tensor eigenvalue is directly related to the entanglement of a state. In this paper we will explore a new connection, which relates the \textit{smallest} tensor eigenvalue to the entanglement of a pure or mixed state. This originates in the fact that the entanglement of a state is related to the positive-definiteness of a tensor, which in turn is linked to the sign of its smallest tensor eigenvalue.

In the present paper we report results of our investigations on the connection between spectral properties of the tensor of order $2j$ associated with a spin-$j$ state, and the classicality of that state. The paper is organised as follows. First we recall some definitions of quantumness and the tensor representation, and show how the spectrum of the tensor is connected to the quantumness/classicality question. In section \ref{SectionTensorEigevalues} we introduce tensor eigenvalues, and as an illustration calculate them explicitly for two examples. In section \ref{SectionCalculatingquantumness} we introduce an efficient algorithm for calculating the distance from a state to the set of classical states. Section \ref{SectionConnectionTensorQuantum} explores numerically the connection between smallest tensor eigenvalue and quantumness.

\section{Definitions}
\label{sec:definitions}

\subsection{Entanglement and quantumness }

We consider a system of $N$ qubits and we restrict the Hilbert space to the subspace of symmetric states. We will describe them with the terminology of spin-$j$ states with $N=2j$. Spin coherent states can be written as \cite{Haake2000} 
\begin{equation}
\label{spin coherent}
\ket{\alpha} =\!\!\! \sum_{m=-j}^j \sqrt{\binom{2j}{j+m}} \left(\cos\frac{\theta}{2}\right)^{j+m}\left(\sin\frac{\theta}{2}e^{-i\phi}\right)^{j-m}\!\!\!\!\!\! \ket{j,m},
\end{equation}
with $\theta \in [0,\pi]$ and $\phi \in [0,2\pi[$ spherical angles. Here $\ket{j,m}$ are the usual angular momentum basis vectors, i.e.~the simultaneous eigenvectors of the total angular momentum squared $\bJ^2$ and its $J_z$ component, with eigenvalues $j(j+1)$ and $m$, respectively ($\hbar=1$). The spin coherent state $\ket{\alpha}$ can be seen as a spin-$j$ pointing in the direction $\bn=(\sin\theta\cos\phi,\sin\theta\sin\phi,\cos\theta)$. A spin-$j$ state $\rho_c$ is classical if and only if it can be expressed as a mixture of spin coherent states with positive weights \cite{GirBraBra08}, i.e.~if there exist spin coherent states $\ket{\alpha_i}$ such that  
\begin{equation}
\label{classicalstates}
\rho_c= \sum_i w_i \ket{\alpha_i}\bra{\alpha_i},\qquad 0 \leq w_i \leq 1,\quad \sum_i w_i =1.
\end{equation}
We denote by $\mathcal{C}$ the ensemble of such states. Since a coherent spin state is formally exactly a pure symmetric separable state and vice versa \cite{etaPaper}, an entangled symmetric multi-qubit state is therefore a state which cannot be written as a classical  state as in Eq.~\eqref{classicalstates}. The amount of entanglement translates into a certain amount of non-classicality, or quantumness, defined as the (Hilbert-Schmidt) distance to the convex set of classical states,
i.e. 
\begin{equation}
 \label{quantumness}
Q(\rho)=\min_{\rho_c\in\mathcal{C}} ||\rho-\rho_c||,
\end{equation}
where $||A|| = \sqrt{\tr (A^\dagger A)}$ is the Hilbert-Schmidt norm
\cite{QQQ}. A spin-$j$ state has a quantumness larger than zero whenever the corresponding $N$-qubit state is entangled. 

It is known that the separable state closest to a symmetric state in terms of the Bures distance is also symmetric \cite{HubKle09}. However for other distances this may not be the case. In particular, the Hilbert-Schmidt distance from an $N$-qubit symmetric state to the set of separable states is in general not equal to the quantumness of the corresponding state of a physical spin-$j$ system, as some separable non-symmetric states may lie closer.

\subsection{Tensor representation}
In order to conveniently deal with expansions of quantum states over spin coherent states, we use a representation suited to this purpose, that has recently
been introduced in \cite{PRL2015}. We express a spin-$j$ density
matrix $\rho$ in the following way. Let $\sigma_{a}$, $1\leq a\leq 3$,
be the usual Pauli matrices, and $\sigma_0$ the $2\times 2$ identity
matrix. We define the $4^{N}$ matrices $S_{\mu_1\dots\mu_{N}}$ (with
$N=2j$) by  
\begin{equation}
 \label{basisS}
S_{\mu_1\ldots\mu_{N}}=P \left(\sigma_{\mu_1} \otimes\sigma_{\mu_2} \cdots \otimes \sigma_{\mu_N}\right) P^\dagger,\quad 0\leq\mu_i\leq 3,
\end{equation}
with $P$ the projector onto the symmetric subspace of tensor products of $N$ spins-$\frac12$ (the subspace spanned by Dicke states). The matrix $\rho$ can be expanded over the $S_{\mu_1\dots\mu_{N}}$ as
\begin{equation}
\rho =\frac{1}{2^{N}}\,A_{\mu_1\mu_2\ldots\mu_{N}}S_{\mu_1\mu_2\ldots\mu_{N}}
\label{canonrhoj}
\end{equation}
(summation over repeated indices is implicit), with real coefficients
\begin{equation}
\label{defcoor}
A_{\mu_1\mu_2\ldots\mu_{N}}=\tr(\rho\, S_{\mu_1\mu_2\ldots\mu_{N}})
\end{equation}
(see \cite{PRL2015} for detail). The $A_{\mu_1\mu_2\ldots\mu_{N}}$ are invariant under permutation of the indices and enjoy the property that for any $\mu_i$, $3\leq i \leq N$ and $0\leq\mu_i\leq 3$,
\begin{equation}
\label{contraction}
\sum_{a=1}^{3}A_{aa\mu_3\ldots\mu_{N}}=A_{00\mu_3\ldots\mu_{N}}.
\end{equation}
Normalization of the states $\rho$ in \eqref{canonrhoj}, $\tr \rho=1$, translates to $A_{00\ldots 0}=1$.

The coordinates $A_{\mu_1\mu_2\ldots\mu_{N}}$ can be seen as a symmetric order-$N$ tensor. We thus refer to \eqref{defcoor} as the tensor representation of $\rho$. This representation is a generalization of the spin-$\frac12$ Bloch sphere representation
\begin{equation}
\rho=\frac{1}{2}\,A_{\mu} S_{\mu}
\label{canonrhoj12}
\end{equation}
with Bloch vector ${\bf A}=\tr(\rho\,{\boldsymbol{\sigma}})$ and $A_0=1$ (noting that $S_{\mu}=\sigma_{\mu}$).

\subsection{Classicality in the tensor representation}
The tensor associated with a spin coherent state $\ket{\alpha}$ pointing in direction $\vec{n}$ is simply given by 
\begin{equation}
\label{TensorCoherentState}
A_{\mu_1\mu_2\ldots\mu_{N}}=\bra{\alpha} S_{\mu_1 \mu_2 \dots \mu_{N}} \ket{\alpha}=n_{\mu_1} n_{\mu_2} \cdots n_{\mu_{N}},
\end{equation} 
with $n_0=1$ and $\bn=(n_1,n_2,n_3)$ \cite{PRL2015}. The definition of classicality, Eq.~\eqref{classicalstates}, can be reexpressed in terms of tensors. A state is classical if and only if there exist positive weights $w_i$ and unit vectors $\vec{n}^{(i)}$ such that its tensor of coordinates $A$ can be written as
\begin{equation}
\label{tensorClassState}
A_{\mu_1 \mu_2 \cdots \mu_{N}}=\sum_i w_i n^{(i)}_{\mu_1} n^{(i)}_{\mu_2} \cdots n^{(i)}_{\mu_{N}},
\end{equation}
with $n_\mu^{(i)}=(1,\vec{n}^{(i)})$.
Contracting such a tensor with an arbitrary real order-1 tensor $q$ gives
\begin{equation}
\label{TensorMustBePositive}
A_{\mu_1 \mu_2 \cdots \mu_{N}} q_{\mu_1} q_{\mu_2}\cdots q_{\mu_{N}}=\sum_i w_i \left(n^{(i)}_\mu q_{\mu} \right)^{N}. 
\end{equation}
If $j$ is an integer (i.e.~if $N$ is even), the right-hand side is
always positive since the weights $w_i$ are positive. Therefore, any
tensor having the form \eqref{tensorClassState} is such that its
contraction with an arbitrary order-1 tensor
is positive. This precisely corresponds to the definition of
positive semi-definiteness of the tensor $A$ as introduced in \cite{Qi05}. A necessary condition for
classicality of $\rho$ is thus that its associated tensor be positive
semi-definite. In the case of a spin-1 system, where the tensor reduces to a matrix, this is also a sufficient condition \cite{Spin1QuantPaper}. However, for $j \geq 2$ it is
not sufficient anymore, since there exist non-classical states which
have a positive tensor representation, as will be discussed below. 

Before continuing the discussion on the relationship between classicality and tensor properties, we introduce some elements of the spectral theory of tensors.

\section{Tensor eigenvalues}
\label{SectionTensorEigevalues}

\subsection{Definitions}
Let $A_{\mu_1 \dots \mu_{N}} $ be the tensor representation of a spin-$j$ state. Its entries are real and symmetric under any permutation of indices. Tensor eigenvalues and eigenvectors of such a real symmetric tensor are defined in \cite{Qi05}. Different definitions have been introduced. For instance, for a tensor with $N$ indices, each ranging from $0$ to $n-1$ (in our case $n=4$), Z-eigenvalues, which we will use in this paper, are the real numbers $\lambda$ such that there exists a real vector $v$ with $n$ components verifying
\begin{align}
\begin{split}
\label{Zeigenvalues1}
A v^{[N-1]}=\lambda v\\
v^T v=1,
\end{split}
\end{align}
where $A v^{[k]}$ denotes the tensor of order $N-k$ given by
\begin{equation}
\left(A v^{[k]}\right)_{\mu_{k+1}\dots\mu_N}=A_{\mu_1 \mu_2 \ldots \mu_N}v_{\mu_1} v_{\mu_2} \cdots v_{\mu_{k}},
\end{equation}
and $v^T$ is the transpose of $v$.

The different definitions of tensor eigenvalues can be written as special cases of the B-eigenvalues, which are defined \cite{CuiDaiNie14} as
\begin{equation}
\label{Beigenvalues}
A v^{[N-1]}=\lambda B v^{[m-1]}, \quad Bv^{[m]}=1,
\end{equation}
where $B$ is a real symmetric order-$m$ tensor and $\lambda, v_{\mu} \in
\mathbb{C}$. If $B$ is chosen as the identity matrix (i.e.~$m=2$) and $\lambda, v_{\mu}$ are restricted to real values, then the solutions $\lambda$ are the Z-eigenvalues defined in Eq.~\eqref{Zeigenvalues1}. If $m=N$ and $B$ is the identity tensor (i.e.~$B_{\mu_1\ldots\mu_n}=1$ if all $\mu_i$ are identical and
$B_{\mu_1\ldots\mu_n}=0$ otherwise), so that $Bx^{[m]}=x_0^{m}+x_1^{m}+\cdots x_n^{m}$, real solutions to \eqref{Beigenvalues} are called H-eigenvalues \cite{Qi05}. Another type are the D-eigenvalues, which have recently found application in magnetic resonance imaging studies of the diffusion kurtosis coefficients of water molecules \cite{Qi08DEigenvalues}. They can be written as real B-eigenvalues if $m=2$ and there exist a symmetric positive definite matrix $D \in \mathbb{R}^{n\times n}$ with $Bx^{[2]}=x^T D x$, such that there exists a real vector $v$ with 
\begin{equation}
Av^{[m-1]}=\lambda D v, \quad v^T Dv=1.
\end{equation} 
For a more detailed overview on the topic of tensor eigenvalues see \cite{TensorOverviewKolda,Qi05, RoughVersionQi}.

It is possible, via resultant theory, to generalize the usual matrix notions of determinant and of characteristic polynomial, and to obtain eigenvalues as the (generally complex) roots of the characteristic polynomial associated with the tensor \cite{Qi05}. Note however that the Z (or H)-eigenvalues defined above are real numbers. If this restriction to reals is lifted, many properties of ordinary matrix eigenvalues are recovered (for instance the number of eigenvalues, or their total sum, is known). Nevertheless, the restriction to real numbers is justified if one wants to generalize the property that a matrix is positive semi-definite if and only if its eigenvalues are positive. Indeed, both Z and H-eigenvalues share the property that a tensor is positive semi-definite if and only if all Z or H-eigenvalues are positive, which makes them the most natural suitable generalization of matrix eigenvalues. But the H-eigenvalues are not invariant under rotation, while Z-eigenvalues are, as will be shown below. Since spin coherent states behave in a very simple way under rotation, we will concentrate on the Z-eigenvalues defined by Eq.~\eqref{Zeigenvalues1}, which we will refer to, from now on, as "tensor eigenvalues". Note that we also tested our methods on the H-eigenvalues, and they gave comparable results to the ones presented in section \ref{SectionConnectionTensorQuantum}.

\subsection{Properties}

Tensor eigenvalues do not share all the properties of the familiar matrix eigenvalues. For example it is in general not true that the tensor eigenvalues of a diagonal tensor are just its diagonal elements. However, the tensor eigenvalues are invariant under rotations and the corresponding eigenvectors are just the rotated eigenvectors (Theorem 7.~of \cite{Qi05}). In order to familiarize the reader with the tensor notation, let us show this explicitly. Take $v$ as a tensor eigenvector of the real symmetric tensor $A$ with tensor eigenvalue $\lambda$, i.e.~fulfilling \eqref{Zeigenvalues1}. Given a real orthogonal matrix $R$ and the rotated objects marked with primes, then 
\begin{align}
&A' v'^{[N-1]}= \prod_{i=1}^N R_{\mu_i,\nu_i}A_{\nu_1 \dots \nu_N} \prod_{j=1}^{N-1} R_{\mu_j,\eta_j} v_{\eta_j}\\
&=\prod_{j=1}^{N-1} (R^TR)_{\nu_j,\eta_j} R_{\mu_N,\nu_N} A_{\nu_1 \dots \nu_N} v_{\eta_j} \\
&= R_{\mu_N,\nu_N}  A_{\nu_1 \dots \nu_N}\prod_{j=1}^{N-1} v_{\nu_j} \stackrel{\eqref{Zeigenvalues1}}{=}  R_{\mu_N,\nu_N} \lambda v_{\nu_N}=\lambda v',
\end{align}
which proves that the eigenvalues are unchanged by rotations and the new eigenvectors are just the rotated old ones. This feature is particularity important in our case, because a rotated spin-$j$ quantum state $\rho'=\hat{R}^\dagger \rho \hat{R}$, with $\hat{R}=\exp (-i\theta\, \vec{J} \cdot \vec{n})$ the spin-$j$ representation of a rotation, has a tensor representation given by $A'_{\mu_1 \dots \mu_N}=R_{\mu_1,\nu_1}\cdots R_{\mu_N,\nu_N}A_{\nu_1\dots\nu_N}$ with $R$ the $4\times 4$ matrix whose $3\times 3$ lower-right block is the orthogonal matrix associated with the rotation of axis $\vec{n}$ and angle $\theta$, and $R_{\mu, 0}=R_{0,\mu}=\delta_{0,\mu}$ \cite{PRL2015}. 
  
Determining tensor eigenvalues is usually a computationally hard problem. It can be expressed in the following way: The tensor eigenvalues defined by \eqref{Zeigenvalues1} are the critical points of the polynomial
\begin{equation}
L(\lambda;x_{1}, x_{2},\ldots,x_{N})=Ax^{[N]}-\lambda\left(||{\bf x}||_2^N-1\right),
\end{equation}
with 
\begin{equation}
||{\bf x}||_2=\sqrt{x_0^2+x_1^2+x_2^2+x_3^2}.
\end{equation}
Indeed, critical points of $L$ are defined by $\nabla L=0$; the conditions $\partial L/\partial x_{\nu}=0$ are equivalent to the first line in Eq.~\eqref{Zeigenvalues1}, as can easily be seen from the fact that if $A$ is a symmetric tensor one has
\begin{equation}
\frac{\partial}{\partial x_{\nu}}Ax^{[N]}=N\left(A x^{[N-1]}\right)_{\nu}
\end{equation}
and
\begin{equation}
\frac{\partial}{\partial x_{\nu}}||{\bf x}||_2^N=N(x_0^2+x_1^2+x_2^2+x_3^2)^{N/2-1}x_{\nu}.
\end{equation}
Condition $\partial L/\partial \lambda=0$ gives the second line in Eq.~\eqref{Zeigenvalues1}. Thus the tensor eigenvalues can be obtained as the local extrema of $Ax^{[N]}$ over the 3-sphere $x_0^2+x_1^2+x_2^2+x_3^2=1$.

As shown in \cite{Li14}, a real symmetric tensor is positive semi-definite, i.e.~$Ax^{[N]}\geq 0$ for all $x$, if and only if all of its tensor eigenvalues are non-negative. Hence, it is sufficient to calculate the smallest tensor eigenvalue to determine the positivity of the tensor. In particular, a tensor can be positive definite only if the tensor has an even number of indices: Otherwise each tensor eigenpair $(\lambda,v)$ has also a negative counterpart $(-\lambda,-v)$, as can be seen by the definition \eqref{Zeigenvalues1}. Numerically, the smallest tensor eigenvalue is obtained by computing the global minimum of $Ax^{[N]}$ over the 3-sphere. Such a problem can be tackled numerically using methods described e.g.~in \cite{CuiDaiNie14}. In the next section we show examples of quantum states where tensor eigenvalues can be derived analytically.

\subsection{Examples}

\subsubsection{Tensor eigenvalues of spin coherent states}
For a spin-$j$ coherent state with Bloch vector ${\bf n}$ the tensor representation $A_{\mu_1 \dots \mu_N}$ takes the simple form \eqref{TensorCoherentState}. In order to deduce all tensor eigenvalues $\lambda$ and eigenvectors $x_{\mu}$, we have to solve Eq.~\eqref{Zeigenvalues1}, which then reads
\begin{eqnarray}
\label{Anx}
A_{\mu_1 \dots \mu_N}x^{[N-1]}  &=&(n_{\mu_1}x_{\mu_1})\ldots \nonumber
(n_{\mu_{N-1}}x_{\mu_{N-1}})n_{\mu_{N}}\\ 
=\lambda x_{\mu_{N}}, \quad ||{\bf x}||_2&=&1.
\end{eqnarray}
Since the tensor eigenvalues are invariant under rotation, we can, without loss of generality, rotate $\bf n$ to the form $(1,0,0)$. This simplifies Eq.~\eqref{Anx} to 
\begin{equation}
(x_0+x_1)^{N-1} \begin{pmatrix}
1\\1\\0\\0
\end{pmatrix}=\lambda \begin{pmatrix}
x_0\\x_1\\x_2\\x_3
\end{pmatrix},\quad ||{\bf x}||_2=1.
\end{equation}
From the third and fourth line it is visible that there are two solutions $\lambda=0$ or $x_2=x_3=0$. If $\lambda=0$, then $x_0=-x_1$ and $x_2,x_3 $ are arbitrary under the restriction $||{\bf x}||_2=1$. Otherwise, $\lambda=\sqrt{2}^N$ for $N$ even, or $\lambda=\pm \sqrt{2}^N$ for $N$ odd, and $x_0=x_1=\pm 1/\sqrt{2},x_2=x_3=0$.
Thus the tensor eigenvalues of a tensor associated with a coherent spin-$j$ state are $(\pm1)^{N} \, 2^{j}$ and $0$. For integer $j$ we recover the fact that the tensor is positive, as it should since a spin coherent state is classical.

\subsubsection{Tensor eigenvalues of the maximally mixed state}

For the maximally mixed state $\rho_0=\frac{1}{N+1}\mathbb{1}_{N+1}$, the tensor representation is given by
\begin{equation}
\label{coorid}
Ax^{N}=\sum_{k=0}^{\floor{j}}\frac{\binom{N}{2k}}{2k+1}x_0^{2(j-k)} \left({x_1^2+x_2^2+x_3^2}\right)^{k},
\end{equation}
where $\floor{\cdot}$ is the floor function \cite{PRL2015}. 
For vectors $\vec{x}$ constrained by $x_0^2+x_1^2+x_2^2+x_3^2=1$, Eq.~\eqref{coorid} can be rewritten as
\begin{equation}
\label{TensorEigenMixMixedExplicit}
Ax^{N}=\sum_{k=0}^{\floor{j}}\frac{\binom{N}{2k}}{2k+1}x_0^{2(j-k)}(1-x_0^2)^{k}:=g(x_0),
\end{equation}
with $-1 \leq x_0 \leq 1$. If $j$ is an integer, $g(x_0)$ is a sum of positive terms and thus larger than zero. The tensor eigenvalues are local extrema of $Ax^{N}$ on the 3-sphere, or equivalently the local extrema of $g(x_0)$ over the interval $[-1,1]$. The  local extrema on the border of the interval, $|x_{0}|=1$, give a tensor eigenvalue $\lambda=1$. Because $g(x_0)$ is symmetric there is a local extremum at $x_0=0$, which gives the tensor eigenvalue $\lambda=1/(N+1)$. For $j\geq 3$ the function $g(x_0)$ has exactly one extremum in the interval $]0,1[$, which gives a third tensor eigenvalue (see Appendix for a proof). Thus for integer $j$ the tensor associated with the maximally mixed state has three tensor eigenvalues and the minimal tensor eigenvalue is $\lambda_{min}=1/(N+1)$. 

For half integer $j$ there are two tensor eigenvalues on the border of the interval which give $\pm 1$. For $j\geq 5/2$ the function $g(x_0)$ has a maximum in $ ]0,1 [$ (see Appendix for a proof), and since $g(x_0)$ is antisymmetric also a corresponding minimum in $ ]-1,0 [$. Thus the tensor has four tensor eigenvalues.

\section{Calculating quantumness}
\label{SectionCalculatingquantumness}
Our goal is to compare quantumness of a spin-$j$ state as measured by the distance \eqref{quantumness} with spectral properties of the tensor associated with it. In order to compute quantumness \eqref{quantumness} efficiently, the calculation can be rewritten as a quadratic optimization problem, by fixing a large number of spin coherent states in the sum \eqref{classicalstates} and optimizing over the weights $w_i$. This is detailed in Section \ref{section Quadratic algorithm}. However, this does not guarantee to find the global minimum, as the decompositions of the closest classical states may involve spin coherent states which do not belong to the large set chosen.  To improve the accuracy of the estimation we will use the outcome of the quadratic optimization as starting point in a linear optimization routine  detailed in Section \ref{seclinear}.

\subsection{Quadratic algorithm}
\label{section Quadratic algorithm}

The state is written as a $[2(N+1)^2]$-dimensional real vector $\vec{r}$, whose entries
are the real and imaginary entries of its density matrix $\rho$ in the
$\ket{j,m}$ basis (or any other fixed basis). In the same 
way the classical state $\rho_c$ in Eq.~\eqref{classicalstates} is written as $ C \vec{w}$, where $C$
is a $[2(N+1)^2] \times M$ real matrix whose $i$th column is given by the real and
imaginary parts of entries of $\ket{\theta_i,\phi_i}\bra{\theta_i,\phi_i}$
expressed in the same basis as $\rho$, $\bf{w}$
is the vector of weights, and $M$ is the number of spin coherent states used in the
sum of the form of Eq.~\eqref{classicalstates}. The squared quantumness can be written
as 
\begin{equation}
\label{squaredQuantumness}
Q^2(\rho)=\min\limits_{C,\vec{w}} \sum_{i=1}^{2(N+1)^2} \left[r_i-(C\vec{w})_i\right]^2,
\end{equation}
which can be expressed as
\begin{equation}
\label{quadratic algorithm}
Q^2(\rho)
=\min\limits_{C,\vec{w}}\left[\vec{w}^T \left( C^T C \right) \vec{w} -  \left(2\vec{r}^T C\right)\vec{w} +\vec{r}^T \vec{r}\right].
\end{equation}
To approximate the solution to this optimization problem we generate a large set of
$M$ $(\sim 800)$ spin coherent states $\ket{\theta_i,\phi_i}$ that determine a matrix
$C$ and a vector $\vec{c}=\left(\vec{r}^T C\right)$, and solve
\begin{align}
\label{explicitquadraticalgorithm}
\min\limits_{\vec{w}} \vec{w}^T (C^TC) \vec{w} -2 \vec{c}^T \vec{w}, \qquad w_i\geq0 ,
\end{align}
(we removed the constant term $\vec{r}^T \vec{r}$). Note that the entries of $(C^TC)$ are given by
\begin{multline}
(C^TC)_{ik}=|\langle \theta_i,\phi_i | \theta_k,\phi_k\rangle |^2   \\
=4^{-j} [1 + \cos\theta_i \cos\theta_k + \cos(\phi_i-\phi_k)\sin\theta_i\sin\theta_k]^{2j} 
\end{multline}
and that
\begin{equation}
\vec{c}_i=\bra{\theta_i,\phi_i}\rho\ket{\theta_i,\phi_i}.
\end{equation}

The optimization \eqref{explicitquadraticalgorithm} can be performed with the powerful numerical algorithms available, e.g.~the 'interior-point-convex' method
\cite{9780521833783}. It is notable that the size of the quadratic
optimization problem, given by the vector $\vec{c}$ and the matrix $ C^TC $,
does not depend on the spin size $j$, but only on the number of random spin coherent states used. However, for very large values of $j$ $(\sim1000)$ even the one-time calculation of $\vec{c}$ and $ C^T C $ can become computationally expensive.

To improve the outcome it is advantageous to iterate the optimization several times with different sets of spin coherent states. In the subsequent iterations, only the spin coherent states with large weights are kept and additional nearby states are added to the set. The set is then completed with random spin coherent states. After typically $\sim 8$ iterations, we take the best outcome as an approximation of the global minimum of \eqref{squaredQuantumness}. This also provides an approximation $\tilde{\rho_c}$ for the true closest classical state $\rho_c$. By construction, $\tilde{\rho_c}$ is a classical state, so that quantumness is necessarily overestimated, since the distance to any classical state gives an upper bound on the quantumness. To further improve its determination, a linear optimization can then be performed as follows.

\subsection{Linear algorithm}
\label{seclinear}
Suppose we have obtained an approximation $\tilde{\rho_c}$ for the closest classical state $\rho_c$ by running the quadratic algorithm above. If the classical state  $\tilde{\rho_c}$ is not exactly on the border of the classical domain, it is possible to move it in the direction of the state $\rho$ while remaining in the classical domain. This yields a better approximation of the global minimum, and thus of the actual quantumness. This step can be formulated as a linear optimization problem by parametrizing the states inbetween the classical state $\tilde{\rho}_c$ and $\rho$, as
\begin{equation}
\rho_k=(1-k)\tilde{\rho}_c +k\rho=\tilde{\rho}_c +k (\rho-\tilde{\rho}_c),
\end{equation}
with $k \in [0,1]$. Now the optimization task is to maximize $k$ under the constraint that $\rho_k$ stays classical, which can be formulated in the form of linear constraints as
\begin{equation}
\sum_i w_i \ket{\theta_i,\phi_i}\bra{\theta_i,\phi_i}+ k(\tilde{\rho}_c-\rho)=\tilde{\rho}_c,
\end{equation}
and the optimization is now performed on $w_i$ and $k$ with $0 \leq w_i \leq 1$, and $k>0$ while $\ket{\theta_i,\phi_i}$ are (a large number of) fixed spin coherent states. Similarly as in section \ref{section Quadratic algorithm}, this optimization problem can be written as
\begin{equation}
\max\limits_{\vec{w},k}\,  k, \quad \mbox{with}\quad C\vec{w} + \left(\boldsymbol{\tilde{r}_c}-\boldsymbol{r}\right) k=\boldsymbol{\tilde{r}_c},
\end{equation}
where the $i$-th columns of $C$ are given by the real and imaginary parts of entries of $\ket{\theta_i,\phi_i}\bra{\theta_i,\phi_i}$, and $\boldsymbol{r},\boldsymbol{\tilde{r}_c}$ are the real and imaginary parts of entries of the density matrices $\rho$ and $\tilde{\rho}_c$. Since a linear optimization is much faster than a quadratic optimization, the set of random spin coherent states used to fix the linear constraints can be much larger, e.g.~usually by two orders of magnitude, and still have a runtime comparable to the quadratic optimization. However, in contrast to the quadratic algorithm the computational demands depend on the spin size $j$, since the number of rows in $C$ scales as $\mathcal{O} (j^2)$. In the results presented in the next section this linear optimization step improves the quadratic results usually by an amount smaller than $10^{-4}$. While this improvement is usually negligible, it becomes relevant to estimate quantumness of states close to the boundary of classical states, and to properly identify classical states. 

\section{Connection between tensor eigenvalues and quantumness}
\label{SectionConnectionTensorQuantum}
\subsection{Tensor eigenvalues for entanglement detection}

\begin{figure}[t!]
\begin{center}
\includegraphics[width=0.48\textwidth]{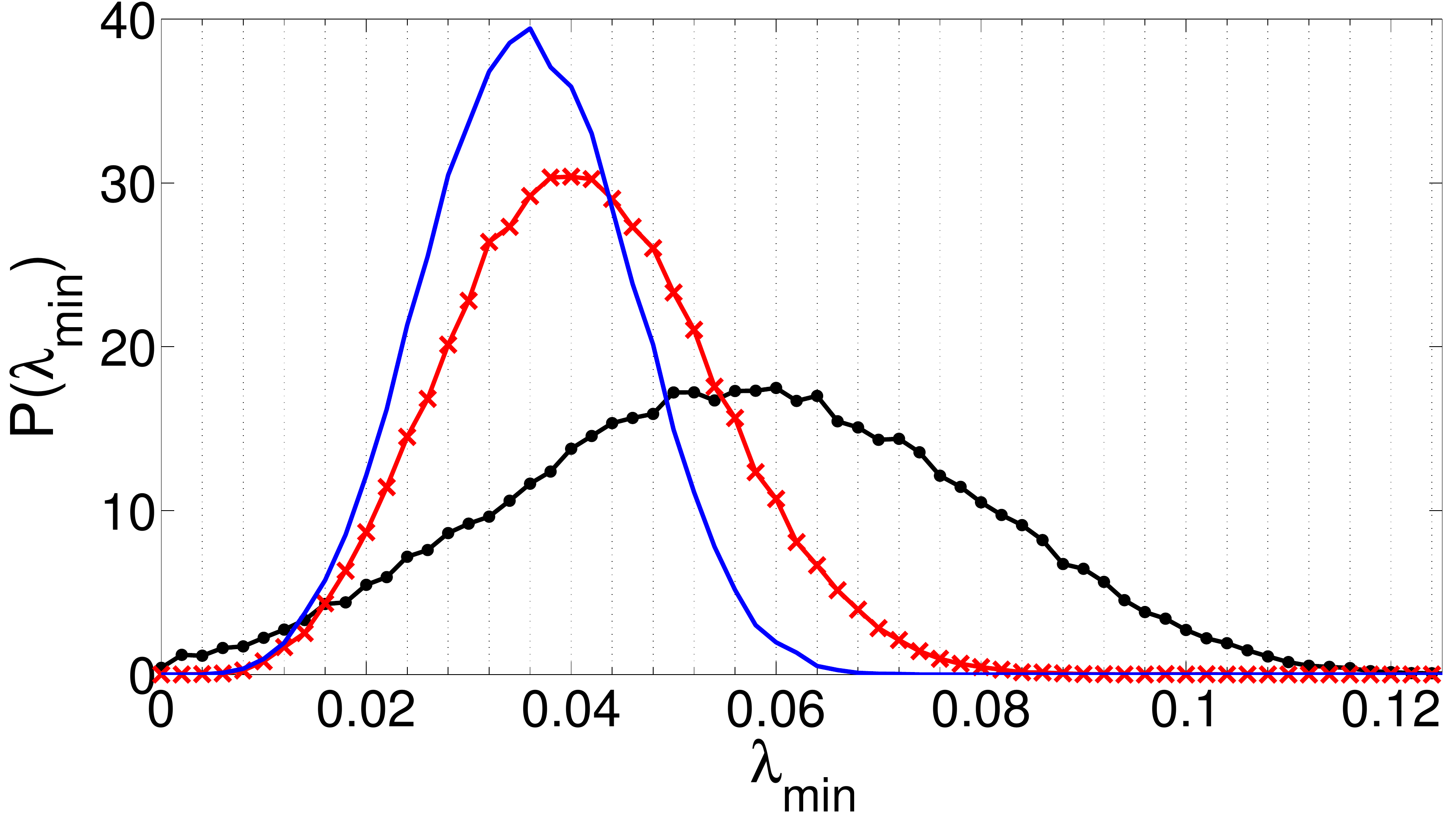}
\end{center}
\caption{(Color online) Probability distribution of the smallest tensor eigenvalue $\lambda_{\min}$ for random states on the border of the classical domain, with $j=2$ (black, dots), $j=3$ (red, crosses) and $j=4$ (blue, solid). These states are the closest classical states to random mixed states and were determined with the quadratic and linear algorithm described in Section \ref{SectionCalculatingquantumness}.} 
\label{fig:histogramZeigOnCCS}
\end{figure}

As mentioned earlier, a classical state must have a positive semi-definite tensor representation. Therefore, if its smallest tensor eigenvalue $\lambda_{\min}$ is negative the state is detected as non-classical, i.e.~entangled. To test the rigour of the detection we generated states just on the border of the set of classical states. This was done by taking random states drawn from the Hilbert-Schmidt ensemble of matrices $\rho=GG^{\dagger}/\tr(GG^{\dagger})$, with $G$ a complex matrix with independent Gaussian entries (see \cite{zyczkowski_generating_2011} for detail), and calculating its closest classical state according to the method presented in the previous section. In Fig.~\ref{fig:histogramZeigOnCCS} the distribution of
the smallest eigenvalues is shown for this ensemble of closest classical states. If positivity of $A$ were a sufficient condition for classicality, then $\lambda_{\min}$ would be equal to 0 for all closest classical states. Numerically, we rather get values centered around $0.03$, $0.04$ and $0.06$ for $j=4,3,2$, respectively. 

Thus, states lying at the border of classical states, with zero quantumness, have a smallest tensor eigenvalue significantly larger than zero, which indicates that for the values of $j$ considered this method of entanglement detection is not well suited for too weakly entangled states.

\begin{figure}[t!]
\begin{center}
\includegraphics[width=0.48\textwidth]{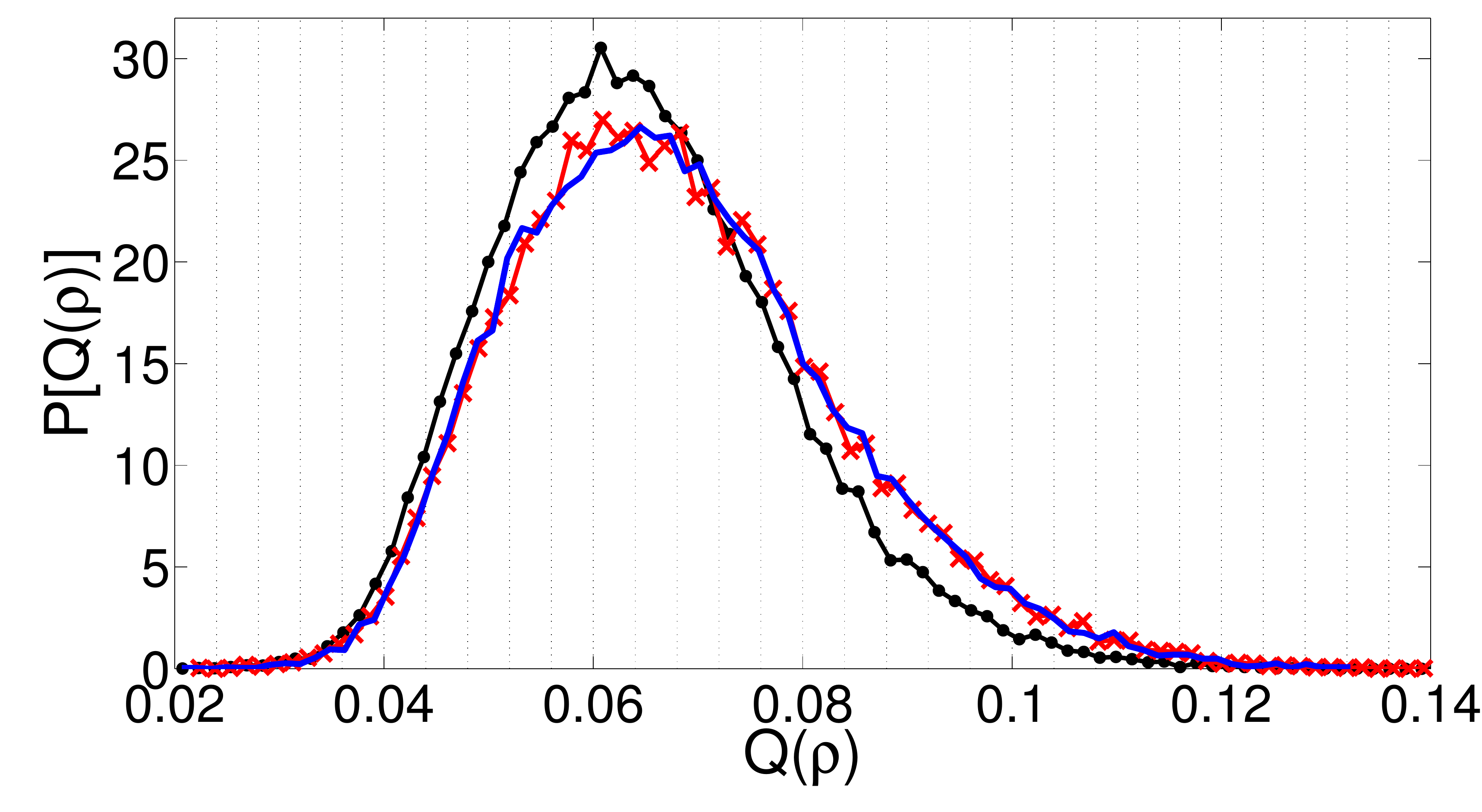}
\end{center}
\caption{(Color online) Probability distribution of the quantumness $Q(\rho)$ \eqref{quantumness} for states having a positive smallest tensor eigenvalue smaller than $10^{-5}$. The states are created by mixing a random mixed state with the maximally mixed state according to \eqref{createStatesIWthZeroEW} and decreasing $a$ until the smallest tensor eigenvalue is close to zero. The numerical uncertainty of the quantumness is of the order $10^{-4}$. The three lines black (dotted), red (crossed), blue (solid) correspond to spin sizes $j=2,3,4$. These states are all entangled, but nevertheless have a positive definite tensor representation.\label{fig:Spin1-3Histogram}}
\end{figure}

Conversely, one may wonder what is the typical quantumness of states which have a vanishing smallest tensor eigenvalue. To investigate this we generated states such that $\lambda_{\min}\simeq 0$, by mixing a random initial state $\rho$ with the maximally mixed state 
\begin{equation}
\label{createStatesIWthZeroEW}
a \rho+(1-a) \frac{1}{N+1}\mathbb{1}, \quad 0\leq a \leq 1
\end{equation}
(with $\mathbb{1}$ the identity matrix), and decreasing $a$ until the smallest tensor eigenvalue was close to zero. The results for these states are shown in Fig.~\ref{fig:Spin1-3Histogram}. The quantumness is distributed around the value of $0.06$, irrespective of the spin size $j$, which again indicates that the smallest tensor eigenvalue is not able to detect weakly entangled states. This appears to be a systematic underperformance, because we did not find instances of classical states which also have a smallest tensor eigenvalue equal to zero. Instead, almost all states on the "detection border" $\lambda_{\min}= 0$ already have a quantumness larger than $0.02$.  

To conclude, the smallest tensor eigenvalue detects entanglement (or quantumness) in spin-2 to spin-4 states only reliably if the quantumness is at least about $0.1$. In the other direction, spin-2 to spin-4 states can be assumed to be separable (or classical) only if the smallest tensor eigenvalue is larger than $0.12$.

\subsection{Measure of entanglement based on tensor eigenvalues}

\begin{figure}[t!]
\begin{center}
\includegraphics[width=0.48\textwidth]{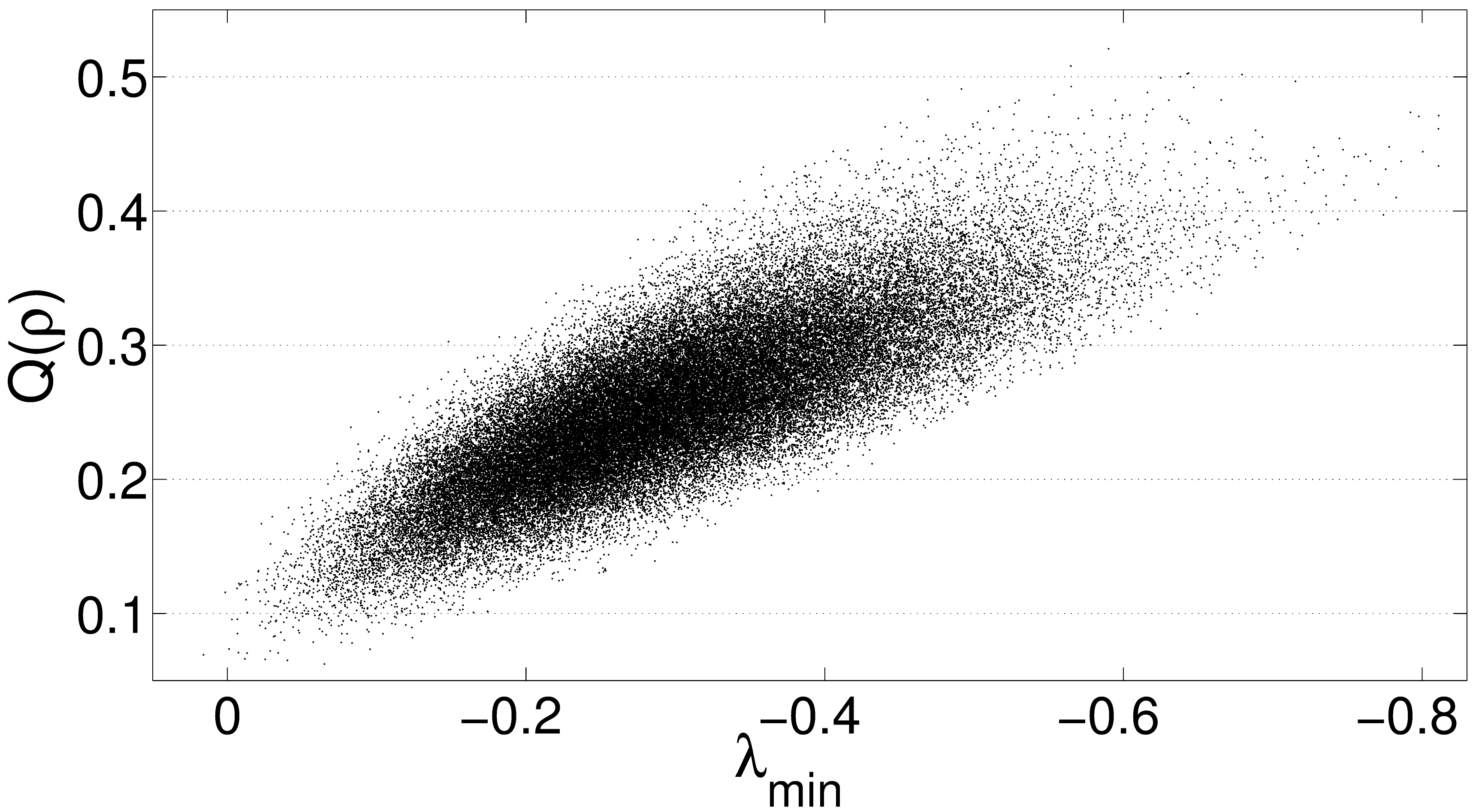}
\end{center}
\begin{center}
\includegraphics[width=0.48\textwidth]{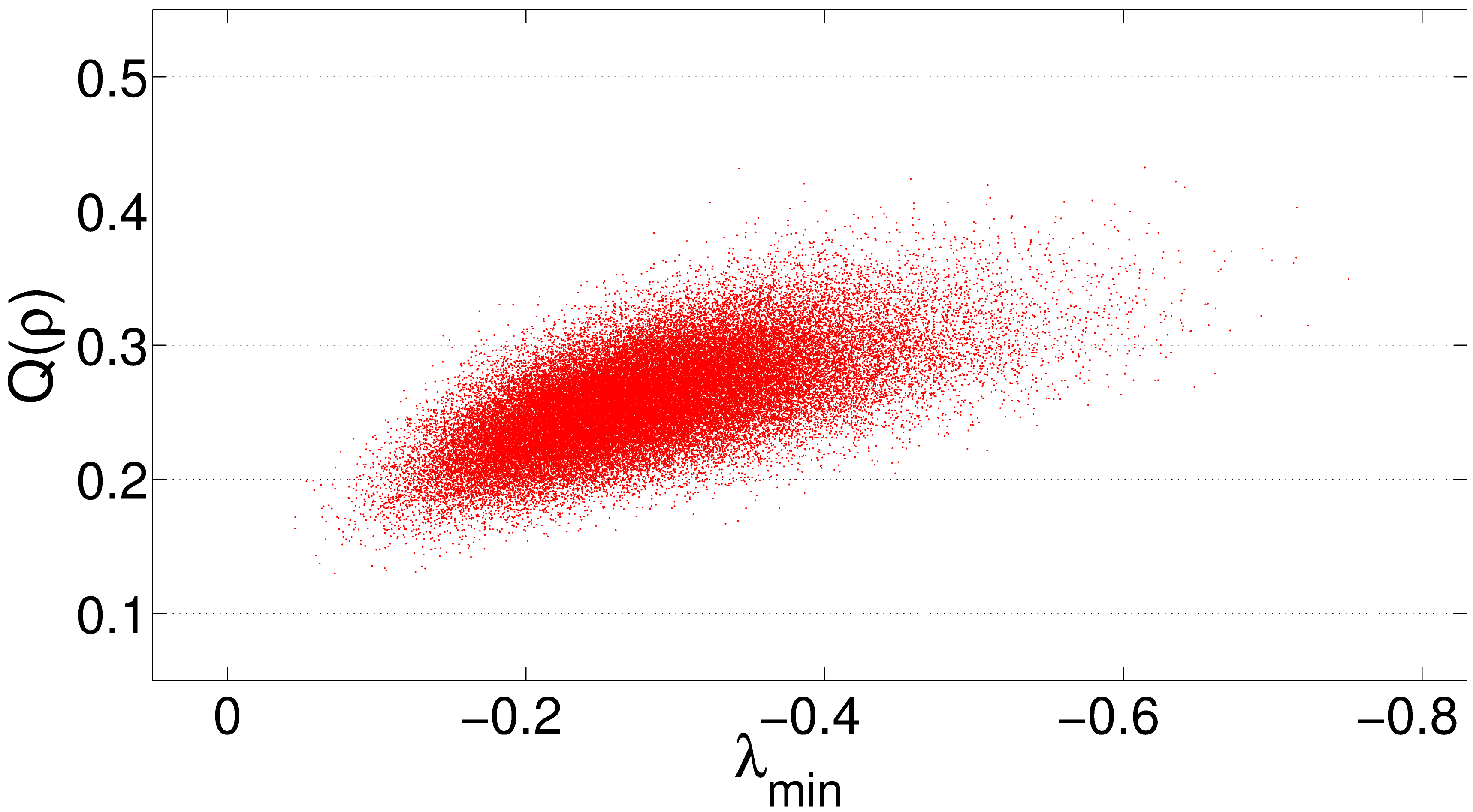}
\end{center}
\begin{center}
\includegraphics[width=0.48\textwidth]{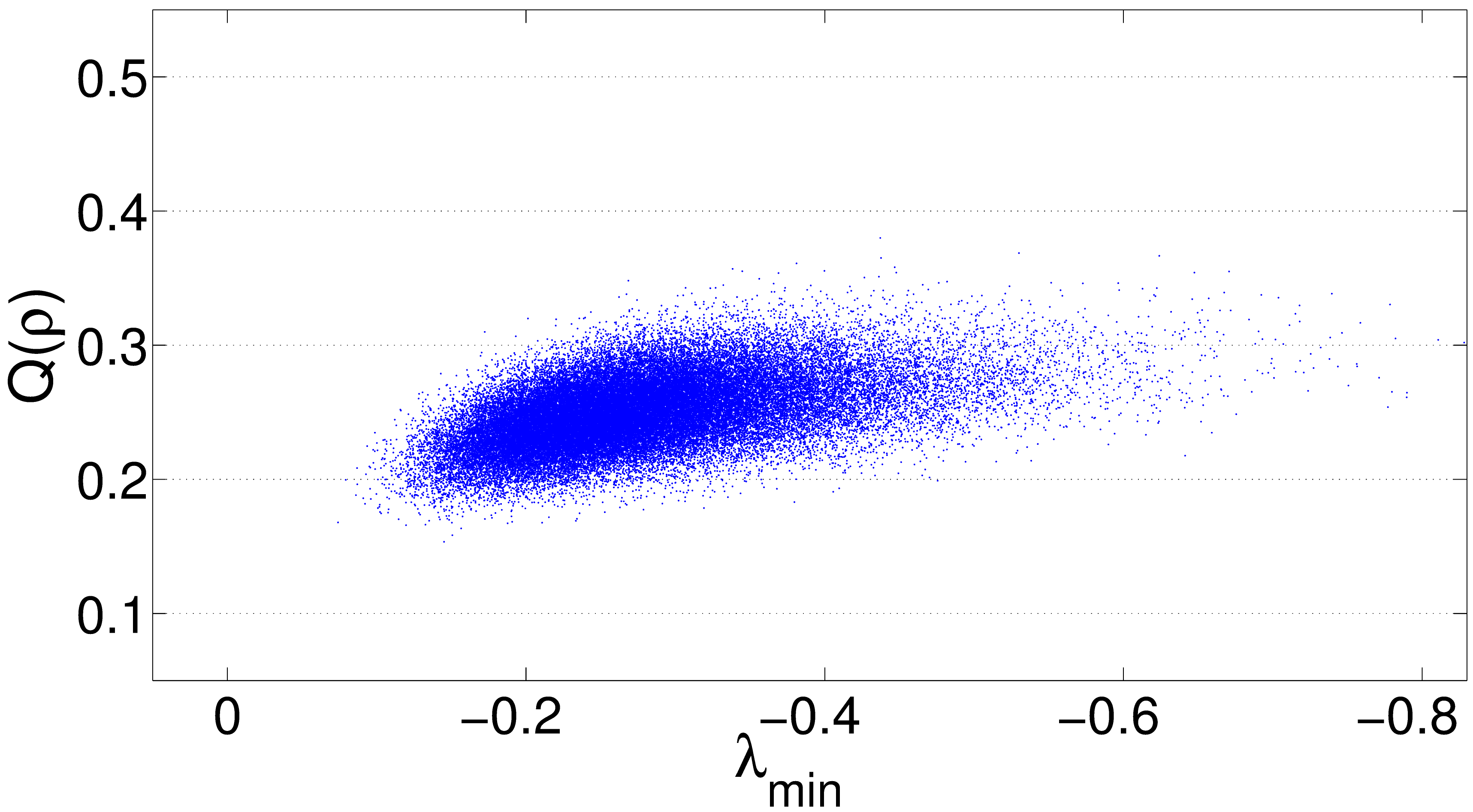}
\end{center}
\caption{(Color online) The quantumness \eqref{quantumness} as function of the smallest tensor eigenvalue \eqref{Zeigenvalues1} for $\sim 60.000$ randomly generated mixed spin-$j$ states. The top figure corresponds to spin size $j=2$, the second to $j=3$ and the bottom to spin size $j=4$. There is a clear correlation between the amount of quantumness and the magnitude of negative smallest tensor eigenvalue, however, this correlation is getting weaker for $j=3$ and even weaker for $j=4$.  } 
\label{fig:spin2-4}
\end{figure}

The results above show that while any state with $\lambda_{\min}<0$ is entangled, positivity of $\lambda_{\min}$ does not seem to be a good indicator of separability. However, for non-classical states, the amount by which $\lambda_{\min}$ is negative is correlated with the amount of entanglement as measured by the quantumness. 

\begin{figure}[t]
\begin{center}
\includegraphics[width=0.48\textwidth]{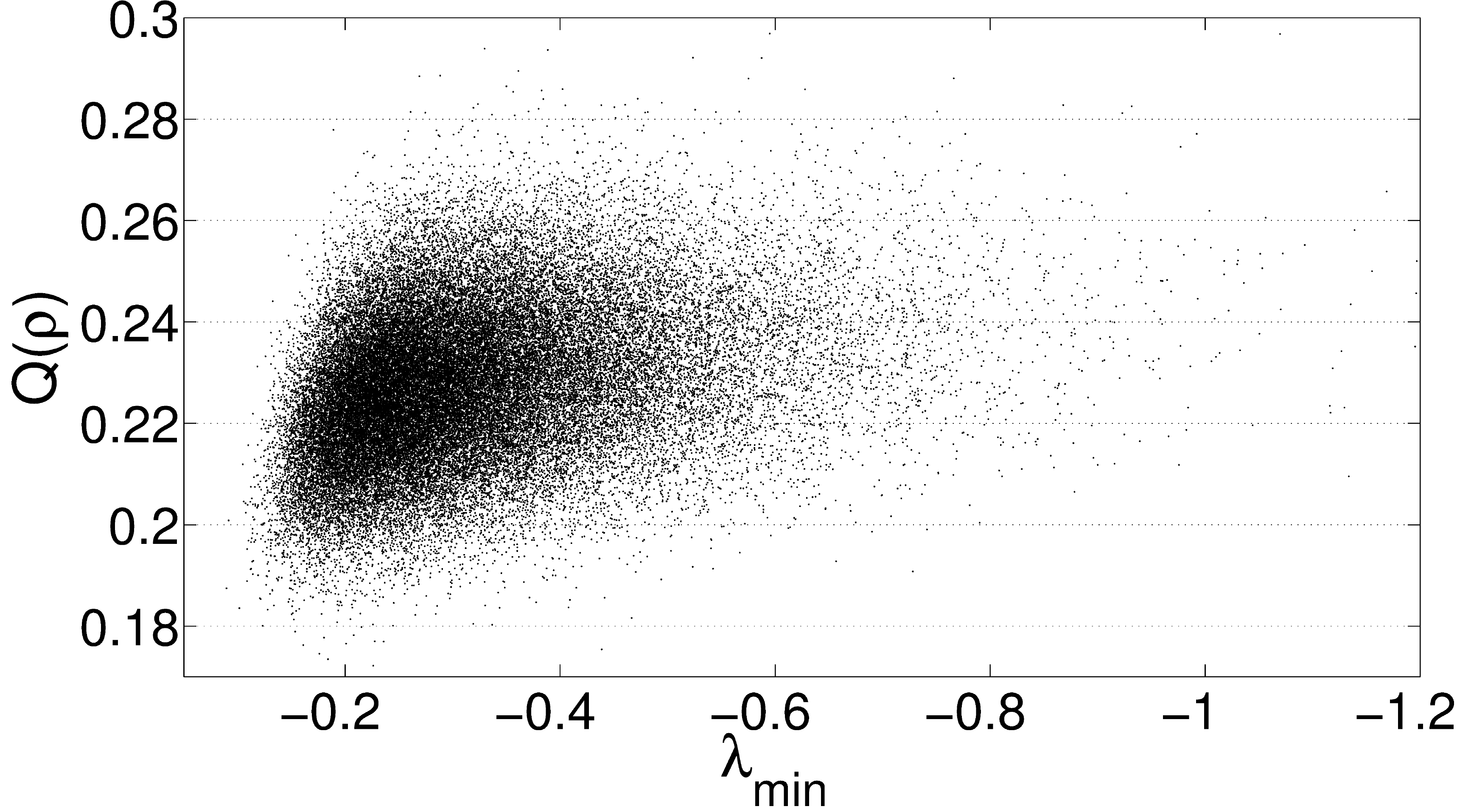}
\end{center}
\caption{(Color Online) The quantumness \eqref{quantumness} of $\sim 60000$ randomly generated spin-6 mixed states as function of their smallest tensor eigenvalue \eqref{Zeigenvalues1}. For this system size there is almost no correlation, between the magnitude of the smallest tensor eigenvalue and the quantumness.} 
\label{fig:spin6}
\end{figure}

This is an approach similar as in the entanglement measure of negativity \cite{negativity}, where the amount of entanglement is taken as the sum of all negative eigenvalues of the partially transposed state $\rho^{\textrm{PT}}$, namely 
\begin{equation}
\label{negativity}
\mathcal{N}(\rho) = \sum_i \frac{|\mu_{i}|-\mu_{i}}{2},
\end{equation}
where $\mu_i$ are the eigenvalues of $\rho^{\textrm{PT}}$. For $j=1$, we showed in \cite{etaPaper} that the tensor eigenvalues are exactly the eigenvalues of $\rho^{\textrm{PT}}$. Unfortunately, in the case of tensor eigenvalues ($j\geq 3/2$), it is computationally expensive to find all tensor eigenvalues. But the smallest tensor eigenvalue provides at least an indicator for the amount of entanglement. This is illustrated in Fig.~\ref{fig:spin2-4}, where quantumness is plotted as a function of the smallest tensor eigenvalue (computed by the algorithms described in Section \ref{SectionCalculatingquantumness}) for a large set of random states. The correlation between the two quantities gets weaker for larger system sizes, i.e.~$j\geq4$. For spin $j=6$, the correlation is almost gone, as can be seen in Fig~\ref{fig:spin6}.

\section{Conclusion}
We introduced a new connection between the mathematical concept of tensor eigenvalues and the
study of entanglement. The smallest tensor eigenvalue can be used to detect
quantumness in symmetric states and can also give an estimator of its
amount. Interestingly, this extends previous results in the mathematical literature relating the largest tensor eigenvalue to the geometric measure of entanglement. For a spin-1, positivity of the smallest tensor eigenvalue is equivalent to separability of the state.  However, for $j\geq 2$ they are not well suited for states which are just 
slightly quantum, since weakly entangled states have usually a
positive semi-definite tensor representation (and are therefore not
detected by the smallest tensor eigenvalue criterion). The correlation between
the amount of quantumness and the magnitude of the (negative) smallest
tensor eigenvalue is noticeable for $j=2,3,4$, but for higher values
of $j$ quantumness and smallest tensor eigenvalues are almost uncorrelated. 
\\

A possible way to improve these results might be to use the sum of all negative tensor eigenvalues as estimator for the quantumness of a state, instead of just the smallest tensor eigenvalue. However, the calculation of all tensor eigenvalue is computationally much more demanding.\\

{\bf Acknowledgments:} We thank the Deutsch-Franz\"osische
Hochschule (Universit\'e franco-allemande) for support, grant
number CT-45-14-II/2015.

\appendix

\section*{Appendix: Tensor eigenvalues of the maximally mixed state}

Here we will prove that the function $g(x)$ defined in \eqref{TensorEigenMixMixedExplicit} has only one local extremum in the open interval $]0,1[$ for $j\geq \frac52$. We reparametrize the function $g$ with 
\begin{align}
x \rightarrow \frac{\cos{t} + \sin{t}}{\sqrt2},
\end{align}
with $t \in ]\frac{\pi}{4},\frac{3 \pi}{4}[$, so that we get  
\begin{align}
g(x) \rightarrow  f(t)=\frac{2^j}{2 j+1} \frac{\cos ^{2 j+1} t - \sin ^{2 j+1} t }{\cos  t  - \sin  t }.
\end{align}
The condition $f'(t)=0$ is equivalent to $H(t)=0$, with 
\begin{multline}
H(t):=(\sin t +\cos t ) \left(\cos^k t  - \sin^k t \right)  \\ + k \sin t    \cos t   (\sin t  -\cos t )   \left(\cos^{k-2} t +\sin^{k-2} t\right),
\end{multline}
with $k=2j+1$. Using $H(\pi/4)=0$ and $H(3\pi/4)\leq 0$, we show that $H(t)$ has only one real root in the interval $]\frac{\pi}{4},\frac{3 \pi}{4}[$ by showing that it is strictly increasing then strictly decreasing then strictly increasing over this interval.

To find the extreme points of $H(t)$ we calculate
\begin{multline}
\label{Hstrich}
H'(t)=(k-1)  (\cos  t -\sin  t  ) \\
\times \left[\sin ^k t-\cos ^k t+k\left(\tan ^2 t \cos ^k t-\frac{\sin ^k t}{\tan  ^2  t}\right)\right].
\end{multline}
Now we show that $H'(t)$ has two roots in $]\frac{\pi}{4},\frac{3 \pi}{4}[$ by setting  $u=\cot t $ in \eqref{Hstrich} with $u \in ]-1,1[$ and counting roots of
\begin{align}
P(u):=-u^k+k u^{k-2}-k u^2 +1
\end{align}
in the interval $]-1,1[$. Descartes' rule of signs tells us that this function has either three or one roots in $]0,\infty[$. As $P(0)=1$, $P(1)=0$, $P'(1)=k(k-5)>0$ and $\lim_{u\to\infty}P(u)=-\infty$, there are necessarily three roots in $]0,\infty[$ and exactly one in $]0,1[$. To study the negative side $u<0$, note that if $k$ is even the function $P(u)$ is symmetric, so that there is also only one root in $u \in ]-1,0[$. In the case of odd $k$, we set $w=-u\in]0,1[$, and 
\begin{align}
P(-u)=\tilde{P}(w)=w^k-kw^{k-2}-k w^2+1.
\end{align}
Applying Descartes' rule again to $\tilde{P}$, we get that $\tilde{P}(w)$ has either two or zero real roots in $]0,\infty[$. However, since $\tilde{P}(0)=1$, $\tilde{P}(1)=2(1-k)<0$ and $\lim_{w\to\infty} \tilde{P}(w)= \infty$, the function has to have exactly one root in the interval $ ]0,1[$ and one in $]1,\infty[$.

This shows that $H'(t)$ has one root in $]\pi/4, \pi/2[$ and one in $]\pi/2, 3\pi/4[$. Since $H'(\pi/2)=1-k<0$, we conclude that $H(t)$ increases, decreases and then increases again, so that it has only one root in $]\pi/4, 3\pi/4[$. So $g(x)$ defined in \eqref{TensorEigenMixMixedExplicit} also has only one extreme point in the open interval $]0,1[$, which gives a single tensor eigenvalue $x\in ]0,1[$.

\end{document}